\documentclass[conference]{IEEEtran}
\IEEEoverridecommandlockouts

\usepackage{amsmath}
\usepackage{amsfonts}
\usepackage{amssymb}

\usepackage{graphicx}
\usepackage{xcolor}

\usepackage[caption=false,font=footnotesize]{subfig}

\usepackage{booktabs}
\usepackage{multirow}
\usepackage{makecell}
\usepackage{diagbox}

\usepackage{algorithm}
\usepackage{algorithmicx}
\usepackage{algpseudocode}

\usepackage{tikz}

\usepackage{enumitem}
\setlist{nolistsep}

\usepackage{url}
\usepackage{balance}
\usepackage{afterpage}
\usepackage[most]{tcolorbox}
\usepackage{subcaption}

\newtcolorbox{mybox}{
enhanced,
boxrule=0pt,frame hidden,
borderline west={4pt}{0pt}{gray!75!black},
colback=gray!10!white,
sharp corners
}

\begin{document}

\title{Towards Energy Efficient Co-Scheduling in HPC}

\author{%
\IEEEauthorblockN{Zhong Zheng}
\IEEEauthorblockA{%
Department of Computer Science \\
University of Illinois Chicago \\
Chicago, USA \\
zzheng33@uic.edu}
\and
\IEEEauthorblockN{Michael E. Papka}
\IEEEauthorblockA{%
Argonne National Laboratory \\
University of Illinois Chicago \\
Lemont, USA \\
papka@anl.edu}
\and
\IEEEauthorblockN{Zhiling Lan}
\IEEEauthorblockA{%
University of Illinois Chicago \\
Argonne National Laboratory \\
Chicago, USA \\
zlan@uic.edu}}

\maketitle

\begin{abstract}
Modern multi-GPU HPC systems expose substantial computational capacity, yet inefficient GPU allocation often leads to wasted energy and underutilization. In practice, GPU applications exhibit heterogeneous and non-linear scaling, making it inefficient to always use all available GPUs. 

We present EcoSched, an online scheduler that jointly optimizes GPU-count selection and application co-scheduling to improve workload-level efficiency on multi-GPU systems. EcoSched uses lightweight runtime profiling to estimate relative performance across GPU counts, applies a score-based policy to balance energy efficiency and idle resources, and incorporates NUMA-aware placement to mitigate interference. We implement EcoSched on heterogeneous CPU–GPU platforms and evaluate it with diverse workloads on H100, A100, and V100 systems. EcoSched achieves up to 14.8\% energy savings, 30.1\% makespan improvement, and 40.4\% EDP reduction over baseline schedulers, with modest performance overhead. These results show that jointly selecting GPU counts and co-scheduling actions is essential for efficient multi-GPU workload execution.

\end{abstract}


\section{Introduction}

High-performance computing (HPC) plays a vital role in advancing numerous research fields by utilizing extensive high-end computational, memory, storage, and network resources to solve complex problems. However, these systems face increasing pressure to deliver higher computational performance while reducing energy use. For example, the exascale Aurora supercomputer at Argonne National Laboratory could incur an annual electricity cost of up to \$30 million when operated at its full 60 MW capacity \cite{Aurora}. Consequently, HPC centers operate under strict facility power limits and rising energy costs. As heterogeneous supercomputers continue to grow in scale and complexity, energy efficiency is now a critical design and operational concern \cite{bergman2008exascale}.


\textbf{Motivation.} As heterogeneous node architectures continue to scale, each compute node integrates more GPUs and exposes substantially greater aggregate compute capacity. However, not every workload can use these resources efficiently. A single-GPU job, for example, leaves the remaining GPUs on a multi-GPU node idle, wasting both compute capacity and idle-GPU energy. Even for multi-GPU jobs, performance scaling is often non-linear: adding more GPUs may yield only marginal runtime improvement because of communication, synchronization, and memory-system overheads. As a result, always assigning a job the maximum available GPUs can be energy-inefficient, since more devices are activated and consume power while delivering limited additional benefit. At the same time, simply reducing GPU counts is not enough, because the released GPUs still consume idle power unless they are reused effectively. Together, these observations suggest that energy-efficient scheduling on multi-GPU nodes requires jointly deciding GPU counts and co-scheduling actions. Because the effective GPU count depends on workload behavior and may vary across applications, such decisions also require lightweight online performance estimation. Here, we define energy efficiency as the total energy required to complete a set of jobs.

\textbf{Limitation of the existing work.} There is considerable prior work on co-scheduling. Cloud-oriented GPU sharing systems have shown that it is valuable to reason jointly about allocation and placement \cite{peng2018optimus, zheng2019cynthia, xiao2018gandiva, gu2019tiresias, park2019accelerated, campos2017scaling}, but they typically rely on elastic execution models such as dynamic scaling, migration, or runtime reconfiguration. These assumptions differ fundamentally from HPC batch environments, where jobs are usually launched with static, quota-based resource allocations through schedulers such as SLURM. Prior HPC co-scheduling work, including Marble \cite{han2020marble}, demonstrates that sharing multi-GPU nodes can improve utilization, but it relies on comprehensive offline profiling and stronger prior knowledge of application behavior. Such assumptions are restrictive in practical scheduling settings, where jobs may arrive from diverse and previously unseen applications. Moreover, Marble generally assumes performance-oriented GPU counts rather than exploiting the system-level energy benefit of selectively downsizing some jobs to create pack-friendly schedules. As a result, prior work still falls short of the coupled online decision problem studied here: jointly choosing GPU counts and co-running job sets to improve energy efficiency under static HPC scheduling constraints.

\textbf{Objective and Challenges.} The observations above point to a coupled scheduling problem: on a multi-GPU node, the scheduler must choose GPU counts and co-scheduling actions jointly in order to minimize energy consumption while keeping performance loss acceptable.

This coupled problem creates three challenges. First, application scaling across GPU counts is heterogeneous and often non-linear, so the energy-efficient operating point is rarely obvious. Second, once GPU-count assignments and co-running job sets must be chosen jointly, the online action space becomes large even for a single node. Third, any such decision must respect interference and placement constraints, because contention in shared CPU, memory, and I/O resources can erase the benefit of otherwise attractive co-scheduling choices.

\textbf{Key insights and contributions.} To address these challenges, we present \textit{EcoSched}, an online energy aware co scheduler for multi GPU nodes.  To address heterogeneous cross-GPU behavior, EcoSched uses lightweight online performance estimation based on brief profiling with GPU DRAM utilization. To address the large coupled action space, it applies a score-based policy that evaluates joint GPU-count and co-scheduling decisions directly. To address interference and placement constraints, it uses NUMA-aware co-allocation to favor packings that remain practical on multi-socket GPU systems.
 
\begin{figure}
    \centering
    \includegraphics[width=1\linewidth]{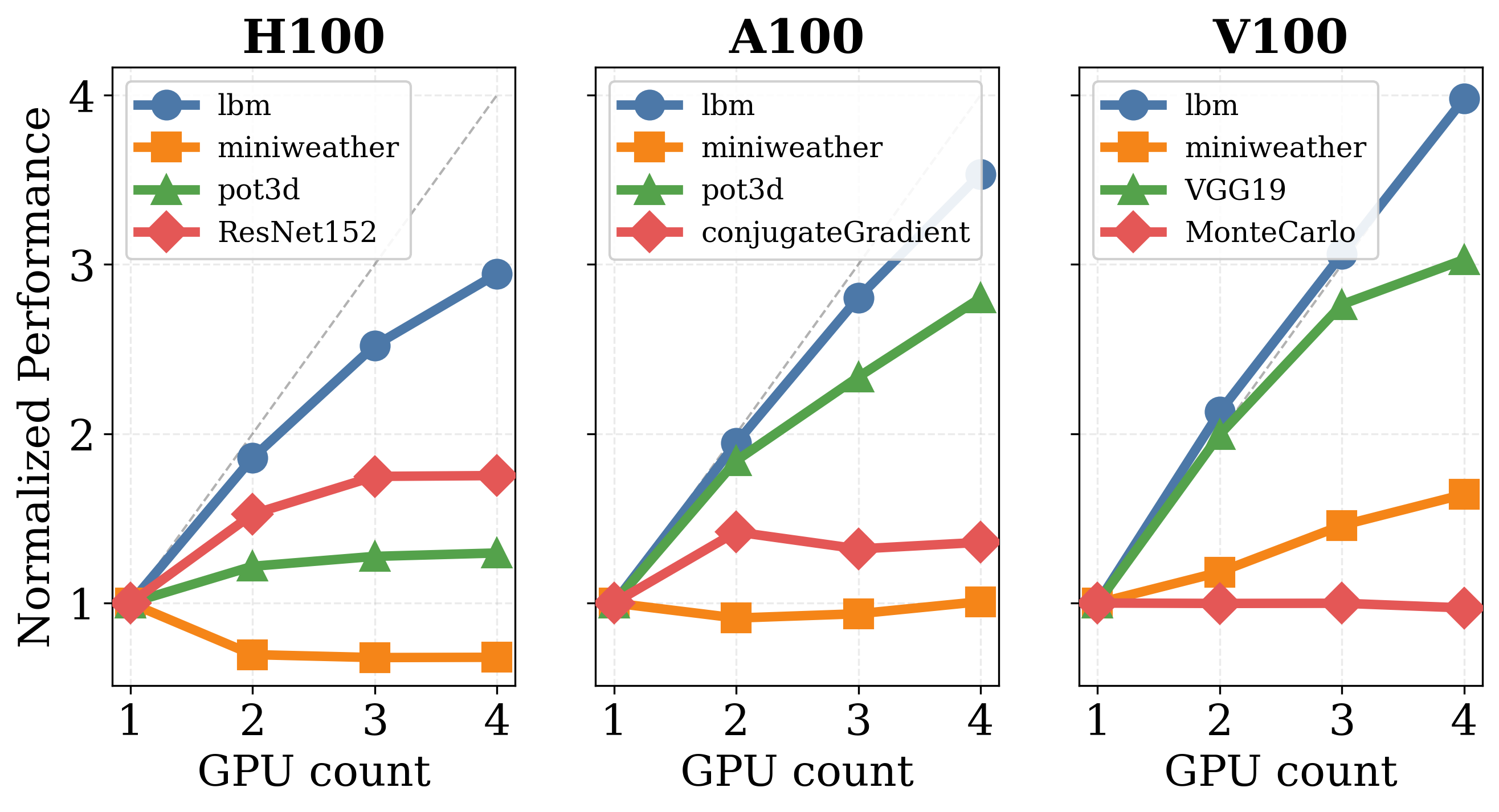}
    \caption{Application performance with different GPU counts. Performance is measured by execution time. Application performance changes are application- and -system dependent.}
    \label{fig:sacling}
\end{figure}

\textbf{Experimental methodology.} We implement EcoSched as an open-source system [link\footnote{link will be provided after acceptance}] and evaluate it with various multi-GPU workloads, from standard benchmarks to real applications, on three multi-GPU platforms (H100, A100, and V100). Experimental results show that EcoSched achieves up to 14.8\% energy savings, 30.1\% makespan improvement, and 40.4\% EDP reduction. Overall, this work offers the following key contributions:
\begin{itemize}
    \item We formulate energy-aware scheduling on multi-GPU nodes as a joint online decision problem over GPU-count selection and co-scheduling under resource and performance constraints.
    \item We design EcoSched, a lightweight online solution that uses runtime signals to estimate relative scaling behavior and guide energy-aware scheduling decisions without exhaustive offline profiling.
    \item We implement EcoSched with NUMA-aware placement and demonstrate end-to-end gains in energy efficiency and workload performance across H100, A100, and V100 platforms.
\end{itemize}

\section{Opportunities and Challenges}\label{sec:motivation}

\begin{figure}
    \centering
    \includegraphics[width=1\linewidth]{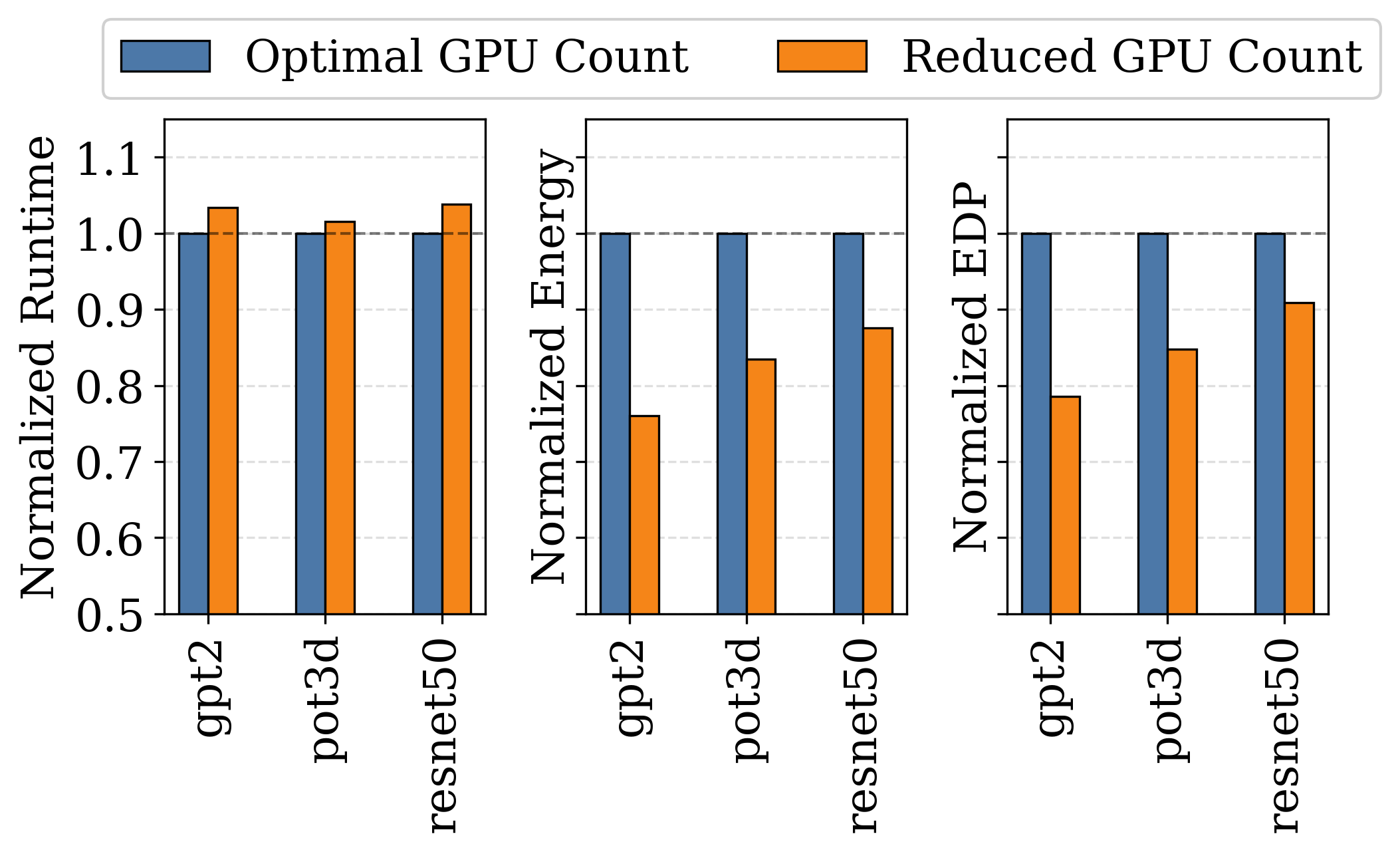}
    \caption{Per-application tradeoff between performance and energy when reducing GPU allocation by one on H100 (\texttt{gpt2} ($3{\rightarrow}2$); \texttt{pot3d} ($4{\rightarrow}3$); \texttt{resnet50} ($4{\rightarrow}3$)). For each application, we compare the performance-optimal GPU count against using one fewer GPU.}
    \label{fig:tradeoff}
\end{figure}

As heterogeneous HPC node architectures continue to scale, each node integrates more GPUs and offers substantially greater aggregate compute capacity. However, in current HPC systems, a node is typically allocated exclusively to one job, and not every job can efficiently use all available GPUs. Some jobs specifically request fewer GPUs than the node can provide, while others exhibit non-linear scaling, where increasing GPU count yields only marginal runtime improvement due to communication, synchronization, and memory-system overheads. Consequently, both case are not resource efficient, as GPU compute capacity and idle-GPU power are wasted. Then, we conduct three motivating experiments that reveal where energy efficiency and node utilization can be improved on a multi-GPU node via co-scheduling. 

\textbf{Opportunity 1: Heterogeneous scaling behavior.} We run representative applications with different GPU counts on H100, A100, and V100 systems; the results are shown in Fig.~\ref{fig:sacling}. Application performance, measured by execution time, improves with additional GPUs in a workload- and system-dependent manner, and the scaling trend is often non-linear. This indicates that assigning the maximum GPU count is not always efficient. For example, the optimal GPU count for \texttt{miniweather} is one on H100, whereas it requires four GPUs on V100 for best performance. These results show that maximum GPU allocation can be inefficient and that the effective GPU count depends strongly on both application behavior and platform characteristics. Therefore, no fixed policy can consistently produce the best allocation across workloads and systems.

\textbf{Opportunity 2: Energy--performance tradeoff.} Next, we run several applications on H100 in two modes: (i) the performance-optimal GPU count and (ii) one fewer GPU than that optimum. We then compare runtime, active energy, and Energy-Delay Product (EDP) \cite{EDP} between the two modes, as shown in Fig.~\ref{fig:tradeoff}. Across applications, reducing GPU count incurs only marginal performance loss but can unlock substantial energy savings. For example, \texttt{GPT2} shows about 3\% performance loss while saving up to 24\% energy. These results show that the GPU count with the best standalone performance is often not the one with the best energy-performance tradeoff. More importantly, this tradeoff is system relevant: accepting a small slowdown for one application can reduce active energy and release GPUs that can be used by other jobs. Thus, selecting GPU counts solely based on solo execution performance can be suboptimal from the perspective of overall system efficiency.

\begin{figure}
    \centering
    \includegraphics[width=1\linewidth]{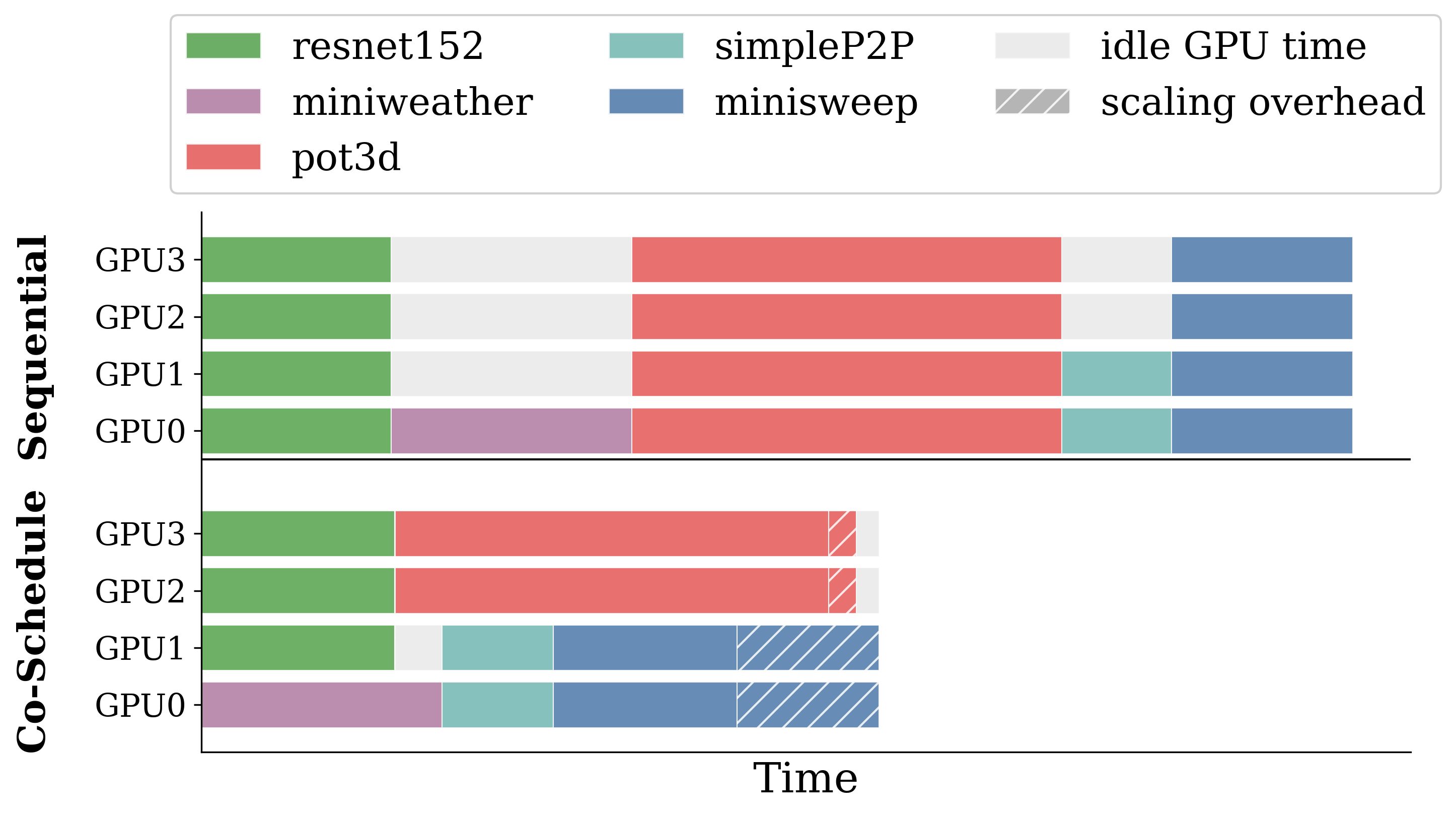}
    \caption{Comparison of scheduling schemes. Experiments are conducted on a 4xH100 system with a two-socket Intel Xeon platform. Sequential: each application runs exclusively using its performance-optimal GPU count. Co-scheduling: multiple applications run concurrently to improve node utilization and energy efficiency.}
    \label{fig:scheduling}
\end{figure}

\textbf{Unlock the opportunities: Co-scheduling.} The first two observations suggest that some applications should run with fewer than their maximum or performance-optimal GPU counts. However, this alone does not fully realize the energy benefit, because the released GPUs still consume idle power if they remain unused. To convert saved GPU capacity into system-level benefit, the scheduler must use the freed GPUs to co-schedule additional applications. Modern HPC CPU platforms, such as Intel Xeon and AMD EPYC, typically use NUMA designs, which provide natural resource partitions for low-interference co-scheduling.

In Fig.~\ref{fig:scheduling}, we compare sequential execution using per-application performance-optimal GPU counts against co-scheduling. The results show that different GPU-count and ordering decisions for the same application queue can substantially reduce idle-GPU time and shorten makespan. For example, assigning two GPUs to \texttt{pot3d} instead of four allows it to co-run with \texttt{simpleP2P} and \texttt{minisweep}. Although \texttt{pot3d} incurs some performance loss, this decision reduces both total energy and makespan. Thus, a GPU count that appears suboptimal in isolation can become preferable at the system level when it enables efficient co-scheduling.

Existing HPC co-scheduling approaches such as Marble \cite{han2020marble} improve utilization, but they do not fully exploit this opportunity because they assume performance-oriented GPU counts rather than jointly leveraging reduced GPU allocations for energy-aware co-scheduling.

Taken together, these observations show that energy-efficient scheduling on multi-GPU nodes cannot be decomposed into two independent steps: selecting a GPU count for each application and then deciding how to pack jobs on the node. The best standalone GPU count may waste energy or leave packable capacity unused, while the best co-scheduling decision depends on which reduced GPU-count modes are available. The scheduler must therefore reason jointly about application scaling, energy–performance tradeoffs, and residual resource reuse.

Designing such a scheduler remains challenging because application scaling varies by GPU count, the joint action space is large, and co-running applications may interfere through shared CPU and memory resources. Specifically, we identify the following key challenges:

\begin{enumerate}

\item \emph{Heterogeneous and non-monotonic scaling across GPU counts.} Application performance does not scale uniformly with increasing GPU count and is application- and system-dependent. Therefore, identifying application performance across different GPU counts online with low overhead is challenging without comprehensive offline profiling.

\item \emph{Coupled GPU-count and co-scheduling decisions.} The GPU-count assignment and co-scheduling combination space is inherently large; arbitrary decisions can miss energy-efficiency opportunities, and global optimization is impractical online because exact application runtimes are unknown. The scheduler must jointly decide which application(s) to run together and how many GPUs to assign to each to maximize the energy efficiency and node utilization.

\item \emph{Interference between co-running applications.} Co-running applications on the same node can cause resource contention in shared components such as CPU cores, LLC, and CPU DRAM bandwidth. Such contention can erode or even negate the energy-efficiency gains of co-scheduling unless placements are carefully coordinated.

\end{enumerate}

\section{Design of EcoSched}

\subsection{Key Ideas and Overview}
 
\begin{figure}
    \centering
    \includegraphics[width=1\linewidth]{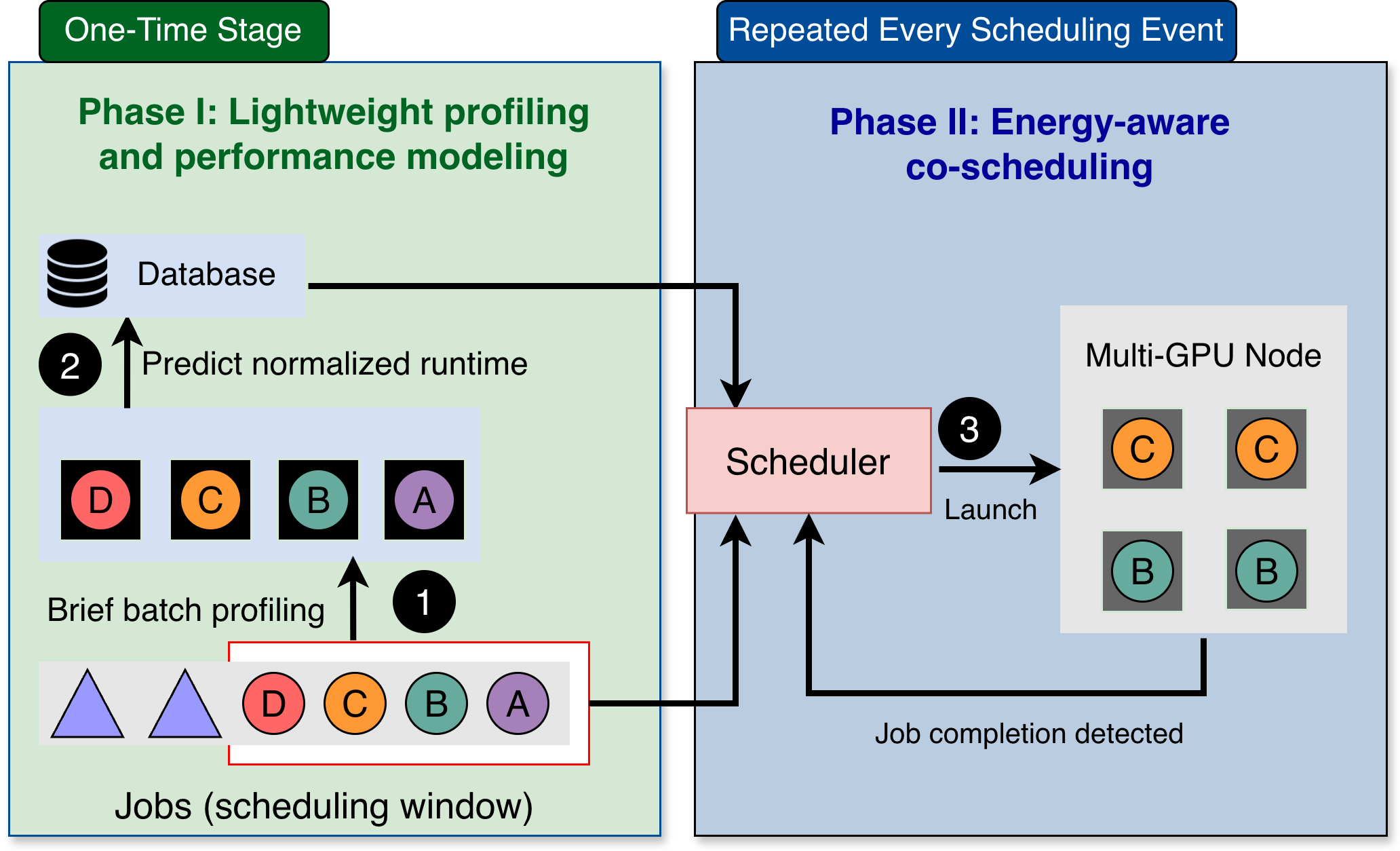}
    \caption{EcoSched consists of two phases: (i) an online performance modeling phase that predicts normalized application runtime under different GPU counts, and (ii) an energy-aware score-based scheduler that co-schedules multiple applications within a scheduling window. }
    \label{fig:workflow}
\end{figure}

As concluded in \S\ref{sec:motivation}, energy-efficient scheduling on multi-GPU nodes is inherently a coupled decision problem: the scheduler must jointly determine GPU counts and co-scheduling actions while accounting for interference. EcoSched addresses this problem through two phases. As shown in Fig.~\ref{fig:workflow}, EcoSched adopts window-based scheduling \cite{window}: instead of dispatching jobs strictly one by one from the head of the queue, it considers a scheduling window formed by a subset of waiting jobs and performs scheduling optimization within that window. Phase~I uses lightweight online profiling based on GPU DRAM utilization to estimate relative application behavior across feasible GPU counts; this phase is performed once for each application in the window and reused throughout subsequent scheduling decisions. Phase~II uses these estimates to make energy-aware co-scheduling decisions under placement constraints via a score-based greedy policy. This design directly maps to the three challenges identified in \S\ref{sec:motivation}: Phase~I addresses lightweight online scaling estimation, Phase~II addresses the joint action-selection problem, and NUMA-aware placement reduces interference among co-running jobs. The following subsections detail both phases.

\begin{figure*}[t]
    \centering

    \subfloat[H100]{%
        \includegraphics[width=0.32\linewidth]{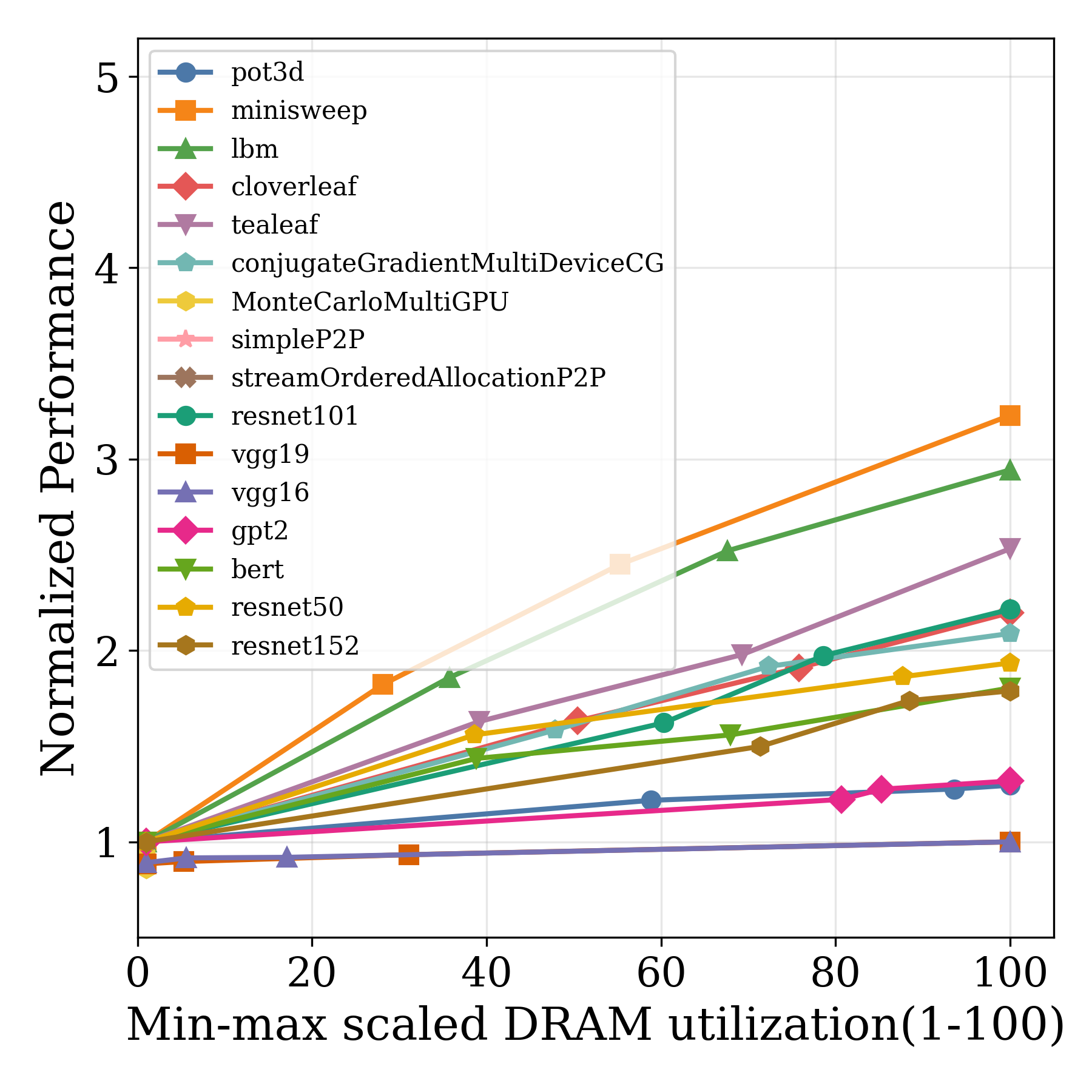}
        \label{fig:dram_perf_h100}
    }\hfill
    \subfloat[A100]{%
        \includegraphics[width=0.32\linewidth]{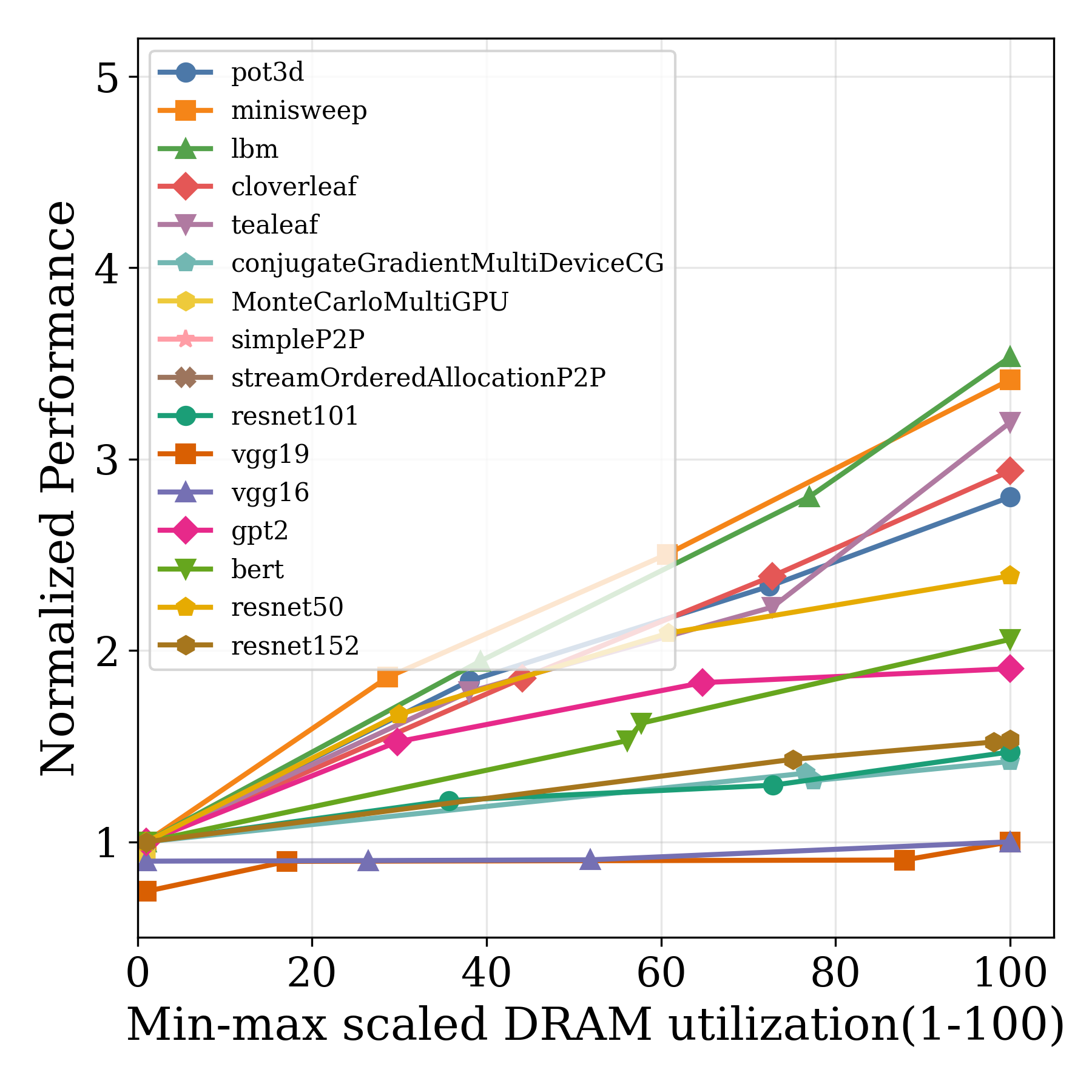}
        \label{fig:dram_perf_a100}
    }\hfill
    \subfloat[V100]{%
        \includegraphics[width=0.32\linewidth]{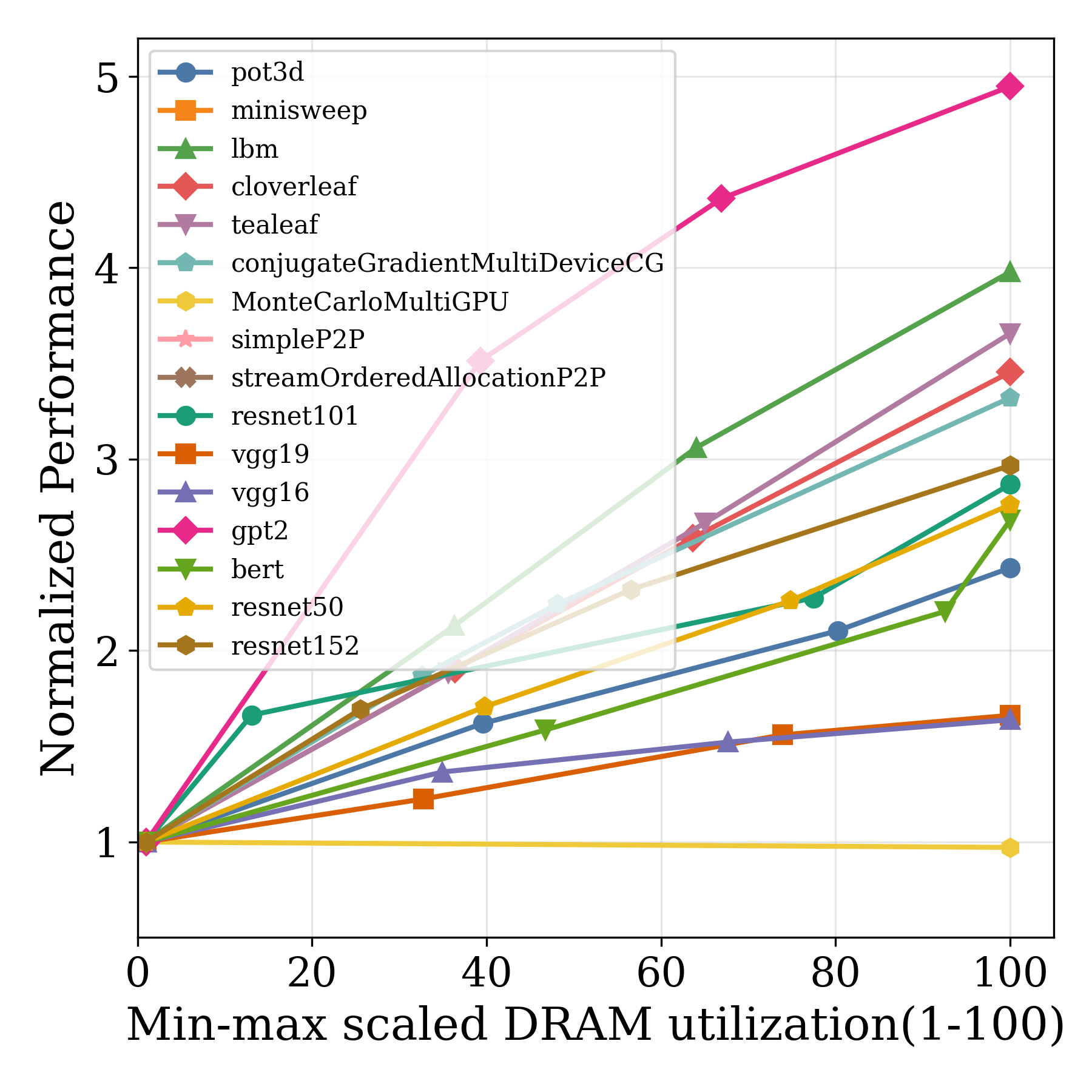}
        \label{fig:dram_perf_v100}
    }

    \caption{Relationship between GPU DRAM utilization and performance (runtime) across H100, A100, and V100 platforms.}
    
    \label{fig:dram_util_vs_perf_all}
\end{figure*}

\subsection{Phase I: Performance Modeling}

To address Challenge~1, Phase~I uses lightweight online profiling to estimate the relative behavior of each application in the current scheduling window across feasible GPU counts. EcoSched does not require exact runtime prediction; instead, it only needs enough relative accuracy to distinguish promising GPU-count modes for scheduling while keeping profiling overhead low.

\textbf{Online Profiling.}
During the online profiling stage, EcoSched builds a lightweight performance model to estimate normalized application performance across different GPU counts. The model is not intended to predict exact runtime values with high precision; rather, it is designed to capture relative scaling behavior across feasible GPU-count modes so that the scheduler can compare them effectively. It briefly profiles each waiting application under feasible GPU-count configurations, and records GPU performance counters together with the corresponding GPU-count settings. These data are then used to construct the model, and this profiling stage only needs to be performed once. EcoSched avoids the cost of exhaustive offline characterization while still capturing the application-specific scaling trends needed for low-overhead online scheduling. Brief profiling is motivated by the fact that many HPC and deep learning workloads exhibit repetitive phase behavior, with recurring patterns over time \cite{online-1}. As a result, short profiling is often sufficient to capture the dominant behavior needed for later scheduling decisions. In typical HPC systems, debugging nodes identical to production nodes are available for short profiling runs; for example, the Aurora supercomputer at Argonne National Laboratory \cite{Aurora} provides 32 debugging nodes. In practice, multiple applications can be profiled simultaneously using these debugging nodes.


\textbf{Performance Prediction.}
To train an effective performance model, selecting an informative yet lightweight signal is essential. Inspired by prior work \cite{online-2}, EcoSched uses GPU DRAM utilization as the primary modeling signal because it captures relative scaling behavior across GPU counts while remaining inexpensive to collect online. EcoSched intentionally avoids building a more complex application-specific model, since its goal is low-overhead online scheduling rather than high-fidelity performance prediction. As shown in Fig.~\ref{fig:dram_util_vs_perf_all}, GPU DRAM utilization is strongly correlated with application performance (measured by runtime).

Building on this observation, EcoSched first uses profiled DRAM-utilization measurements to estimate relative runtime behavior across GPU-count configurations. EcoSched maps these measurements to predicted normalized runtime values, denoted by $\hat{T}^{norm}_{i,g}$, so that GPU-count choices can be compared consistently within each application. Please note that each application has its own performance model.

EcoSched then combines these runtime estimates with profiled power information to obtain a relative energy measure for scheduling. Specifically, it computes the average busy power $\bar{P}_{i,g}$ from online profiling for each feasible GPU count and combines it with the predicted normalized runtime to form a normalized energy proxy, $\tilde{E}_{i,g} = \bar{P}_{i,g} \cdot \hat{T}^{norm}_{i,g}$. This proxy captures the relative energy cost of each application--GPU-count mode and is used in subsequent scheduling decisions.

\subsection{Phase II: Energy-Aware Co-Scheduling}

To address Challenge~2, Phase~II uses the estimates produced in Phase~I to evaluate scheduling actions that jointly specify which applications to launch and how many GPUs to assign to each. This is the mechanism by which EcoSched addresses the coupled decision problem identified in \S\ref{sec:motivation}: rather than deciding GPU counts and co-scheduling separately, it scores joint actions under the current system state and selects accordingly. 

\textbf{NUMA-Aware Resource Partitioning.}
To address Challenge~3, EcoSched uses NUMA-aware resource partitioning to limit interference among co-running applications. Most current supercomputers naturally expose NUMA structure. For example, each Aurora node \cite{Aurora} has two Intel Xeon sockets, and although each Frontier node \cite{frontier} has a single AMD EPYC socket, it still supports NUMA partitioning through NPS \cite{NPS}. EcoSched enforces isolation across co-running applications in terms of CPU cores, LLC, and CPU DRAM bandwidth by assigning different applications to different NUMA domains. On a node with $K$ NUMA domains, we therefore co-allocate at most $K$ applications. This is a deliberate conservative design choice: it gives up some packing flexibility in exchange for lower contention and more predictable energy-performance behavior. GPU allocation, however, need not align exactly with NUMA boundaries. For example, on a 4-GPU node with 2 NUMA domains, one application may be assigned 3 GPUs while another receives 1 GPU, even though their CPU-side resources remain partitioned by NUMA domain. This design is effective because the dominant interference arises from shared CPU cores, LLC, and CPU memory bandwidth, whereas each GPU has its own local memory and PCIe connectivity. By enforcing NUMA locality on the CPU side while allowing flexible GPU-count assignments, EcoSched preserves most of the benefit of co-scheduling while limiting cross-application interference.

\textbf{Score-Based Action Selection.} 
The objective of co-scheduling is to minimize total system energy consumption during workload execution. In our setting, total energy consists of two components: the active energy consumed by executing applications and the energy wasted by GPUs that remain idle while applications run with fewer than the maximum available GPUs. In principle, this objective can be addressed through global optimization over the entire workload schedule. However, such an approach is impractical in online settings because the exact execution time of each application is not known a priori. 

Instead, EcoSched adopts a greedy local decision strategy. At each scheduling event, it evaluates the feasible actions under the current system state and selects the action that minimizes the immediate energy cost, namely the combination of predicted application normalized active energy and idle-GPU energy. We define an \emph{action} \(a\) as a feasible set of application--mode assignments selected at a scheduling decision point. Each mode \(m \in a\) specifies an application and an assigned GPU count. All actions must satisfy current GPU capacity and NUMA placement constraints. EcoSched retains only GPU-count modes whose predicted slowdown is within a tolerance $\tau$ of the best predicted mode. This filtering step removes clearly suboptimal configurations and constrains the scheduling stage to performance-feasible operating points. In this way, EcoSched approximates the global energy-minimization objective through a sequence of local energy-aware scheduling decisions.

The ideal objective is to minimize total energy to completion over the workload. In online settings, however, exact global optimization is impractical because exact runtimes and future system states are unknown. EcoSched therefore uses a local surrogate score as a practical online approximation to this objective. The score captures the two dominant immediate costs of a scheduling decision: selecting application modes with poor energy efficiency and leaving available GPU capacity idle:
\begin{equation}
\begin{aligned}
S(a) &= R_{\mathrm{energy}}(a) + \lambda I(a), \\
R_{\mathrm{energy}}(a) &= \frac{1}{|a|} \sum_{m \in a} \left(\hat{E}^{\mathrm{norm}}_{m} - 1\right), \\
I(a) &= \frac{G_{\mathrm{free}} - G(a)}{M}.
\end{aligned}
\label{eq:score_components}
\end{equation}
Here, \(R_{\mathrm{energy}}(a)\) is the average normalized energy regret of the selected modes in action \(a\), where \(\hat{E}^{\mathrm{norm}}_{m}\) denotes the predicted normalized energy of mode \(m\), and \(|a|\) is the number of modes in \(a\). A value of \(\hat{E}^{\mathrm{norm}}_{m}=1\) corresponds to the best predicted energy configuration for that application. The term \(I(a)\) is the fraction of GPU capacity left idle after launching \(a\), where \(G_{\mathrm{free}}\) is the number of currently available GPUs, \(G(a)\) is the number of GPUs consumed by \(a\), and \(M\) is the total number of GPUs in the node. The weight \(\lambda \ge 0\) controls the tradeoff between minimizing energy regret and reducing idle resources. Finally, EcoSched selects the action with the minimum score among all feasible actions (\(\mathcal{A}_{\mathrm{feas}}\)): 
\begin{equation}
a^{*} = \arg\min_{a \in \mathcal{A}_{\mathrm{feas}}} S(a),
\label{eq:selection}
\end{equation}

\subsection{Putting It All Together}

EcoSched synthesizes the three design elements introduced above into a unified online scheduling framework. Phase~I addresses Challenge~1 by providing lightweight cross-mode estimation of relative runtime and energy behavior across feasible GPU-count configurations. Phase~II addresses Challenge~2 by evaluating joint actions that simultaneously determine which applications to launch and how many GPUs to assign to each. NUMA-aware placement addresses Challenge~3 by constraining co-allocation to reduce interference during execution. Together, these components allow EcoSched to approximate the coupled energy-aware scheduling problem with low online overhead. In operation, EcoSched performs Phase~I once for each application, then repeatedly constructs feasible actions under the current GPU-capacity and NUMA-placement constraints, evaluates them using the score function in Eq.~\ref{eq:score_components}, launches the minimum-score action, and re-invokes the same procedure whenever resources are freed. This iterative loop continues until all applications finish, allowing EcoSched to adapt its decisions to the evolving system state while preserving the low-overhead design goal.



\begin{table}[htbp]
    \centering
    \caption{List of Multi-GPU Workloads.}
    \setlength{\tabcolsep}{2pt} 
    \begin{tabular}{l l l c}
        \toprule
        \textbf{Suite} &  \textbf{App} & \textbf{Input} \\
        \midrule
        \makecell[l]{NVIDIA CUDA\\Code Samples} & \texttt{conjugateGradient} & default \\
        & \texttt{MonteCarlo} & default \\
        & \texttt{simpleP2P} & default \\
        & \texttt{streamOrderedAllocation} & default \\
        \midrule
        \makecell[l]{SPEC HPC\\Benchmarks} & \texttt{lbm} & X=1200 Y=4800 \\
        & \texttt{cloverleaf}  & X=7680, Y=7680 \\
        & \texttt{tealeaf}  & X=10000, Y=10000 \\
        & \texttt{minisweep}  & X=Y=Z=128 \\
        & \texttt{pot3d}  & \makecell[l]{NR=133, NT=361,\\NP=901} \\
        & \texttt{miniweather} & NX=1600, NZ=800 \\
        && BOXES=8 \\
        \midrule
        ML Training & \texttt{ResNet101} & ImageNet \\
        & \texttt{ResNet152}  & ImageNet \\
        & \texttt{ResNet50}  & ImageNet \\
        & \texttt{VGG19}  & ImageNet \\
        & \texttt{VGG16}  & ImageNet \\
        & \texttt{BERT}  & Wikipedia \\
        & \texttt{GPT2}  & Wikipedia \\
        \bottomrule
    \end{tabular}
    \label{tab:apps}
\end{table}

\section{Experimental Setup} \label{setup}

In our experiments, we include two sequential scheduling baselines: \texttt{sequential\_max\_gpu} and \texttt{sequential\_optimal\_gpu}, which respectively assign each multi-GPU application the maximum available GPUs and the GPU count that yields the lowest execution time. We compare EcoSched against these two baselines, a state-of-the-art method (Marble \cite{han2020marble}), and an Oracle policy that assumes perfect prior knowledge. 

We obtain the Oracle schedule by formulating the offline energy-minimization problem as a constraint programming model and solving it with the CP-SAT \cite{cp-sat} solver in Google OR-Tools. Specifically, we model each job as selecting one GPU-count and placement configuration subject to GPU-capacity, NUMA-capacity, and concurrency constraints, while minimizing total node energy, including both application active energy and idle-GPU energy over the makespan. The CP-SAT solver computes the globally optimized schedule offline under the assumption of perfect knowledge of application runtime and power profiles. We then replay the optimized plan to measure the corresponding Oracle execution result and use it as an offline lower-bound baseline for comparison against online scheduling policies.

\textbf{Multi-GPU workloads.} As presented in Table \ref{tab:apps}, we assemble a diverse suite of 17 multi-GPU benchmarks and applications: four from NVIDIA CUDA Benchmarks \cite{cuda_sample}; six from SPEC Benchmarks \cite{dixit1991spec}, a benchmark suite covering representative HPC workloads from diverse scientific domains; seven neural network training workloads. Across our experiments, the workload queues are constructed from this full application pool, so the reported results collectively cover all applications listed in Table \ref{tab:apps}. In our evaluation, this application pool forms a single scheduling window considered by EcoSched. The scheduler's objective is to optimize job selection and GPU allocation within this window. Thus, the current study evaluates optimization within one finite job window rather than a streaming-job setting with continuous arrivals.

\textbf{Multi-GPU systems.} Three different multi-GPU systems are used in our evaluation:

\begin{itemize}
    \item \textit{System 1}: two Intel(R) Xeon(R) Platinum 8468 processors paired with four NVIDIA H100-80GB GPUs.
    \item \textit{System 2}: two Intel(R) Xeon(R) Platinum 8380 processors paired with four NVIDIA A100-80GB GPUs. 
    \item \textit{System 3}: two Intel(R) Xeon(R) Gold 6152 processors paired with four NVIDIA V100-32GB GPU. 
\end{itemize}

We employ NVIDIA NVML \cite{NVML} for GPU power monitoring, and NVIDIA DCGM \cite{DCGM} for monitoring performance counters. Our approach is not limited to NVML and is compatible with any platform that provides equivalent power and performance counter monitoring capabilities. We leverages Linux \texttt{numactl} and \texttt{CUDA\_VISIBLE\_DEVICES} for CPU-core/DRAM partitioning and GPU allocation, respectively.

\textbf{Metrics.} EcoSched’s objective is to minimize total energy to
completion of all workloads while meeting the performance loss bound, hence we
use the following metrics:

\begin{itemize}
    \item \textit{Energy Saving}: percentage reduction in total energy relative to a baseline.
    \item \textit{Makespan Improvement}: the percentage reduction in total completion time.
    \item \textit{EDP Saving}: the percentage reduction in end-to-end EDP.
    \item \textit{Performance Loss}: percentage increase in runtime relative to solo execution with the performance-optimal GPU count.
\end{itemize}

\section{Results}


\subsection{End-to-End Performance}

\begin{figure}[t]
    \centering
    \subfloat[Baseline: sequential\_optimal\_gpu]{%
        \includegraphics[width=1\linewidth]{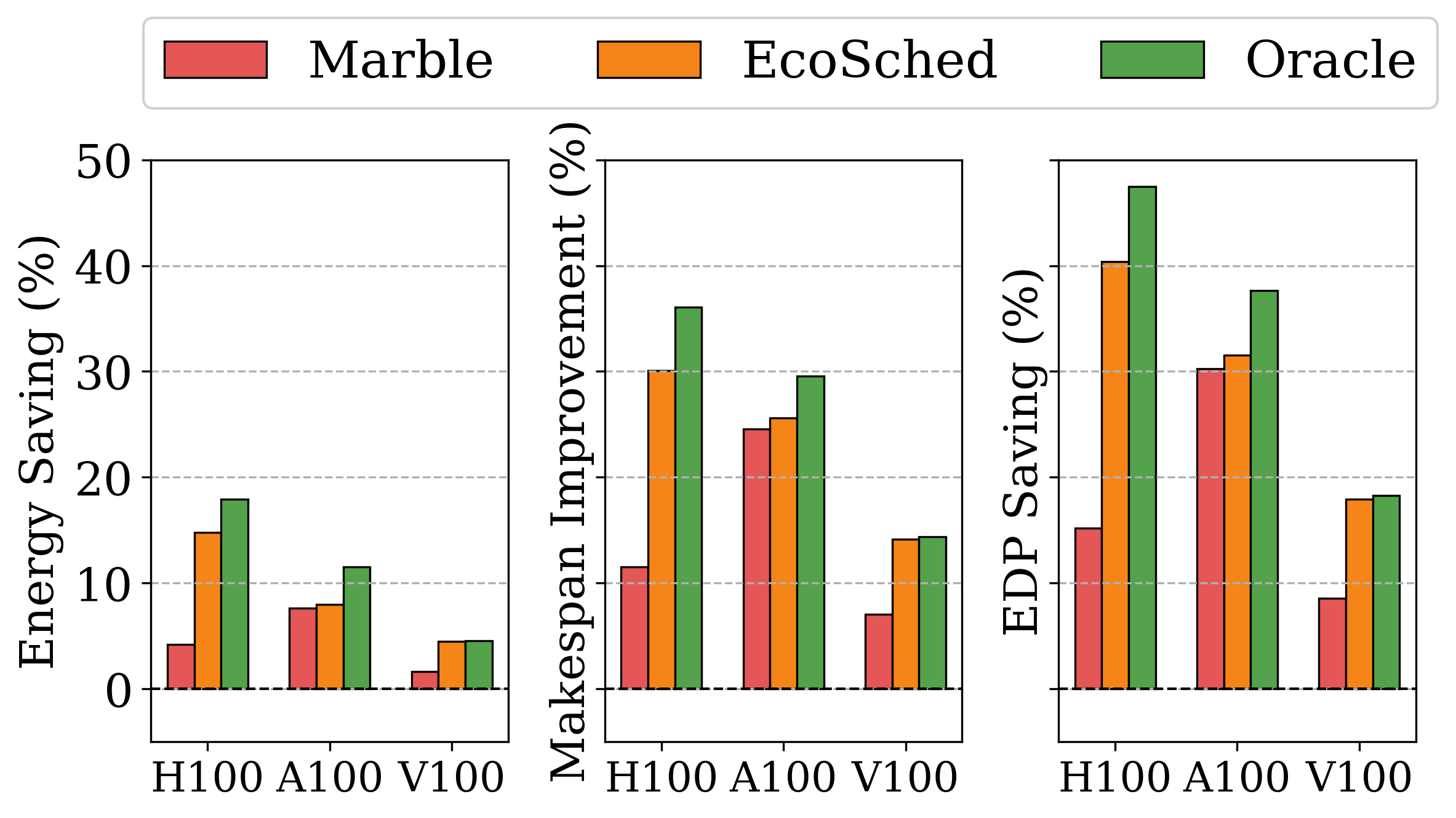}
        \label{fig:end-to-end-opt}
    }\\
    \subfloat[Baseline: sequential\_max\_gpu]{%
        \includegraphics[width=1\linewidth]{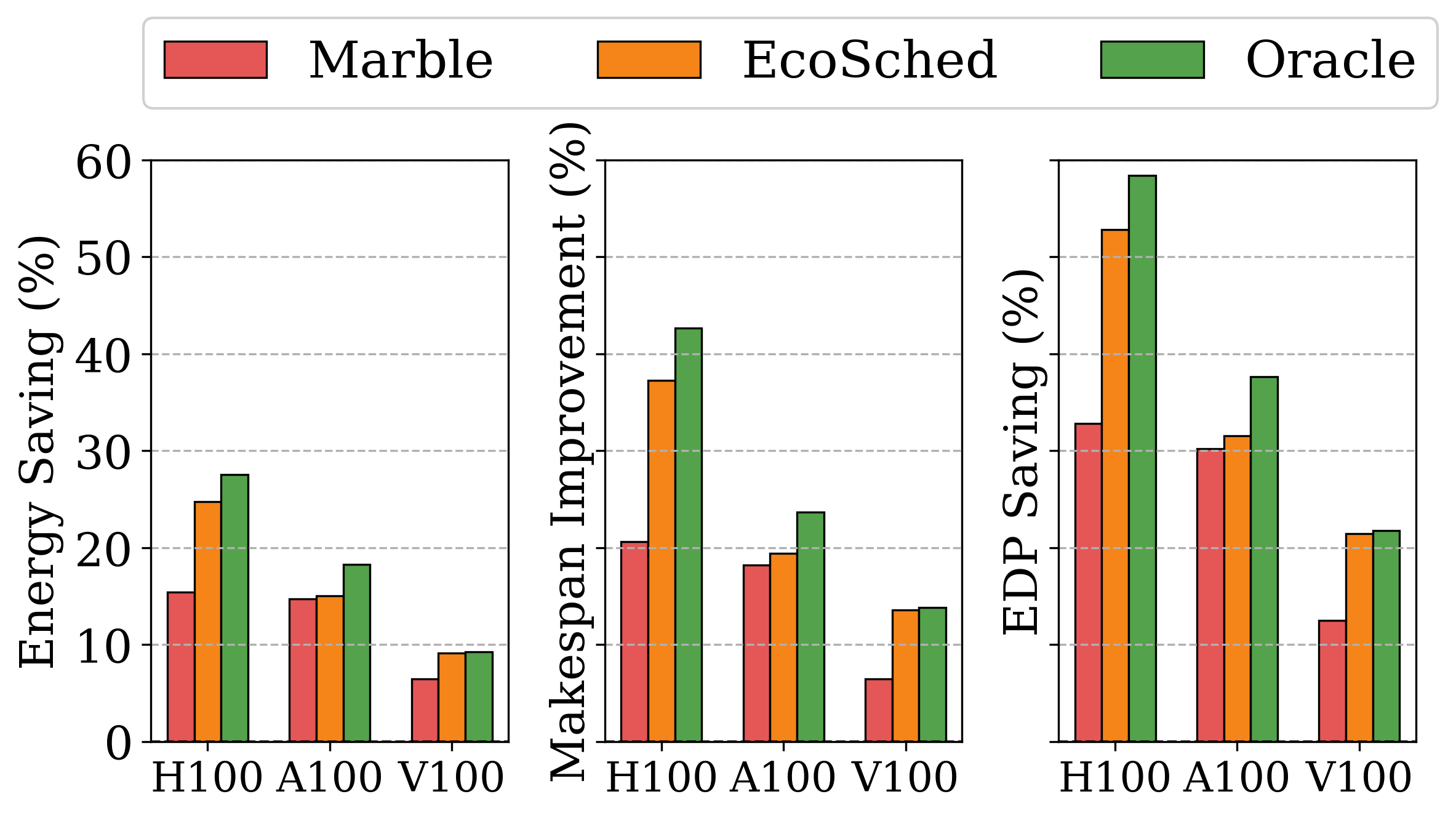}
        \label{fig:end-to-end-max}
    }
    \caption{Scheduling strategy comparison in energy saving, makespan improvement, and EDP saving on H100, A100, and V100 systems under two sequential baselines: sequential\_optimal\_gpu and sequential\_max\_gpu (\S\ref{setup}).}
    \label{fig:end-to-end}
    \vspace{-1.5em}
\end{figure}

This subsection evaluates the central thesis of the paper: jointly choosing GPU counts and co-scheduling actions improves end-to-end system efficiency. Figure~\ref{fig:end-to-end} compares EcoSched against two sequential baselines, Marble, and an offline Oracle on H100, A100, and V100 systems, using end-to-end energy saving, makespan improvement, and EDP saving relative to the \texttt{sequential\_optimal\_gpu} and \texttt{sequential\_max\_gpu} baselines.

We highlight several observations. First, EcoSched improves end-to-end energy, makespan, and EDP relative to the sequential baselines. On H100, EcoSched delivers the largest end-to-end benefit, achieving 14.8\% energy saving, 30.1\% makespan improvement, and 40.4\% EDP saving over the \texttt{sequential\_optimal\_gpu} baseline. EcoSched also narrows much of the gap to the Oracle, which reaches 17.9\% energy saving and 47.5\% EDP saving. Second, EcoSched outperforms Marble because it explicitly exploits the energy--performance trade-off in GPU-count selection: it accepts small per-application slowdowns when doing so unlocks more energy-efficient packing and additional co-scheduling opportunities. On H100, these gains substantially exceed Marble, which improves energy by only 4.2\% and makespan by 11.5\%.

In addition, we observe that all co-scheduling policies, including Marble, EcoSched, and Oracle, show larger improvements over the \texttt{sequential\_max\_gpu} baseline than over the \texttt{sequential\_optimal\_gpu} baseline. This is because some applications achieve their best performance with fewer GPUs, while allocating more GPUs often provides only marginal runtime benefit at substantially higher energy cost.

Third, the cross-platform differences are themselves informative: gains are larger on platforms where more applications exhibit packable slack and smaller on platforms where workloads remain more compute-bound. H100 and A100 provide the greatest opportunity for EcoSched’s joint GPU-count selection and co-scheduling policy to improve both packing and energy efficiency, whereas V100 offers less scheduling slack. This is expected because V100 provides lower compute capacity than H100 and A100; consequently, more applications remain compute-bound on V100 and require more GPUs to achieve their best performance, leaving less room for co-scheduling gains. Even so, EcoSched still achieves 4.4\% energy saving, 14.1\% makespan improvement, and 17.9\% EDP saving, compared with 1.6\%, 7.0\%, and 8.5\% for Marble. Moreover, EcoSched is nearly indistinguishable from the Oracle on V100, where the Oracle achieves 4.5\% energy saving and 18.2\% EDP saving. This result shows that when the available scheduling opportunity is modest, EcoSched can still extract nearly all achievable end-to-end benefit.

The GPU-count choices selected by EcoSched vary substantially across platforms and applications, as shown in Table~\ref{tab:ecosched_gpu_count}. For example, applications such as \texttt{vgg16}, \texttt{vgg19}, \texttt{pot3d}, \texttt{gpt2}, and the ResNet models receive different GPU allocations on H100, A100, and V100, while other applications consistently favor smaller or larger allocations. This variation directly shows why fixed GPU-count policies are inherently limited: the effective GPU count is both application-dependent and platform-dependent, so no single allocation rule can deliver the best energy-performance tradeoff across architectures and workloads. These results therefore motivate EcoSched’s lightweight performance-modeling approach, which selects GPU counts adaptively online rather than relying on a fixed allocation policy or exhaustive offline profiling.

Taken together, these results show that EcoSched consistently improves end-to-end system efficiency across architectures, with particularly large gains on A100 and H100, while remaining close to the offline Oracle. We next examine the mechanism behind this advantage.

\begin{table}[!t]
\centering
\caption{GPU-count choices selected by EcoSched across GPU platforms.}
\label{tab:ecosched_gpu_count}
\scriptsize
\begin{tabular}{lccc}
\toprule
Application & H100 & A100 & V100 \\
\midrule
\texttt{bert} & 4 & 4 & 3 \\
\texttt{cloverleaf} & 4 & 4 & 4 \\
\texttt{conjugateGradient} & 4 & 2 & 4 \\
\texttt{gpt2} & 2 & 4 & 4 \\
\texttt{lbm} & 4 & 4 & 4 \\
\texttt{minisweep} & 4 & 4 & 4 \\
\texttt{miniweather} & 1 & 1 & 1 \\
\texttt{MonteCarlo} & 1 & 1 & 1 \\
\texttt{pot3d} & 2 & 4 & 4 \\
\texttt{resnet101} & 3 & 2 & 3 \\
\texttt{resnet152} & 3 & 2 & 4 \\
\texttt{resnet50} & 3 & 4 & 4 \\
\texttt{simpleP2P} & 2 & 2 & 2 \\
\texttt{streamOrderedP2P} & 2 & 2 & 2 \\
\texttt{tealeaf} & 4 & 4 & 4 \\
\texttt{vgg16} & 1 & 2 & 3 \\
\texttt{vgg19} & 1 & 1 & 4 \\
\bottomrule
\end{tabular}
\end{table}

\subsection{Mechanism Analysis}

\begin{figure}
    \centering
    \includegraphics[width=1\linewidth]{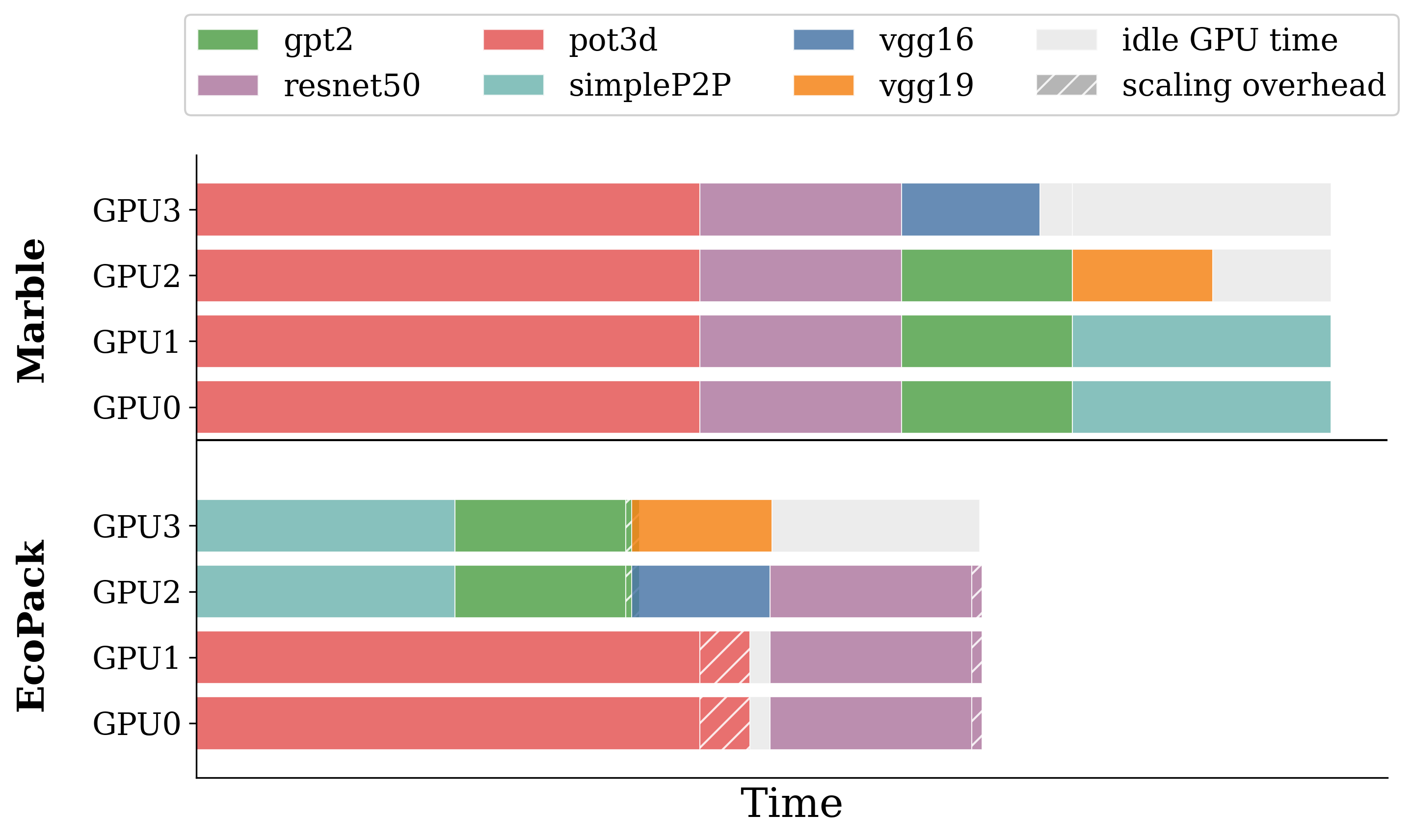}
    \caption{Scheduling six applications on system 1. Marble picks each job's best-performing GPU count, forcing \texttt{pot3d} and \texttt{resnet50} to occupy all 4 GPUs and serializing the workload. EcoSched accepts small scaling overheads to shrink GPU footprints, letting \texttt{pot3d} run concurrently with the other jobs and cutting makespan by 30\%.}
    \label{fig:case_study_scheduling}
\end{figure}

\begin{figure}
    \centering
    \includegraphics[width=1\linewidth]{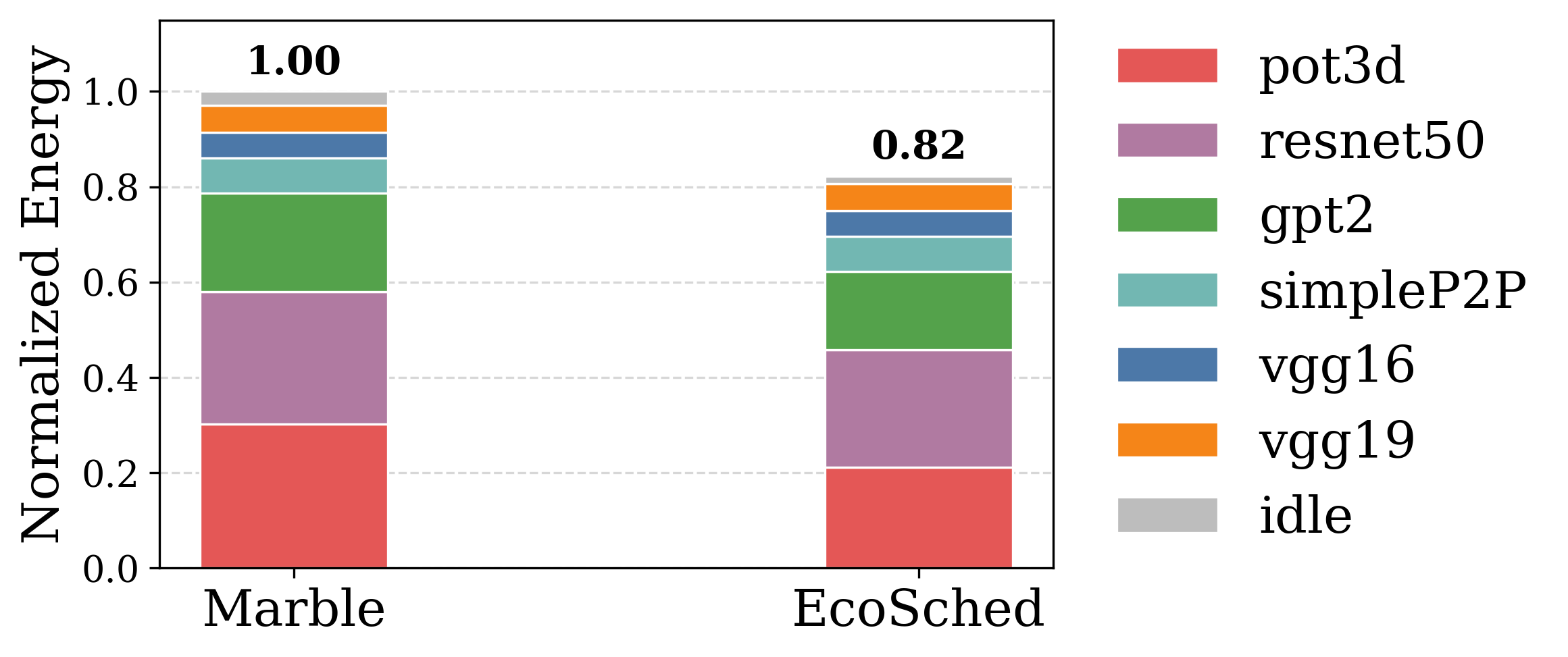}
    \caption{Per-application energy breakdown for the case-study workload, normalized to Marble's total. EcoSched reduces total energy by 17\%.}
    \label{fig:case_study_energy}
    \vspace{-1em}
\end{figure}

To understand why EcoSched works, we analyze the mechanism behind its joint scheduling decisions. The key idea is selective GPU-count downsizing: for applications whose scaling curves flatten early, EcoSched chooses smaller GPU allocations to create more pack-friendly schedules, reducing idle-GPU time and total GPU-seconds. We illustrate this behavior on a representative six-application workload on System 1. Figures~\ref{fig:case_study_scheduling} and~\ref{fig:case_study_energy} show how these decisions reshape execution and lower end-to-end energy.

EcoSched down-sizes the three jobs whose strong-scaling curves flatten earliest: \texttt{pot3d} ($4{\rightarrow}2$ GPUs, $10\%$ slowdown), \texttt{resnet50} ($4{\rightarrow}3$ GPUs, $5\%$ slowdown), and \texttt{gpt2} ($3{\rightarrow}2$ GPUs, $8\%$ slowdown). These small concessions fundamentally change the schedule shape---\texttt{pot3d} now occupies only half the node, opening a wide concurrent lane in which \texttt{simpleP2P}, \texttt{gpt2}, \texttt{vgg16}, and \texttt{vgg19} all execute in parallel with it. \texttt{resnet50} is then squeezed into the brief tail. The hatched regions in Figure~\ref{fig:case_study_energy} make the trade-off visible: the slowdown on each individual job is small, while the parallelism it unlocks is large.

The packed schedule reduces the end-to-end makespan by $\sim$$30\%$ relative to Marble. Crucially, the savings are not paid for in energy: although EcoSched runs the down-sized jobs at slightly higher per-GPU power (running on fewer GPUs concentrates the work), the drop in total GPU-seconds dominates, yielding a $17\%$ reduction in total workload energy (Figure~\ref{fig:case_study_energy}). 

The energy savings are concentrated in the three applications EcoSched reconfigured---\texttt{pot3d}, \texttt{resnet50}, and \texttt{gpt2}---highlighting the value of leveraging EcoSched's energy--performance trade-off. More broadly, this case study shows that EcoSched wins by reshaping workloads into more energy-efficient and pack-friendly execution patterns, rather than optimizing each job in isolation. The best-performing GPU count is often not the most pack-friendly one, and selectively downsizing the right jobs can simultaneously reduce makespan and energy through joint scheduling.

\subsection{Overhead}

Because EcoSched relies on online profiling and event-driven scheduling, we next verify that its runtime and profiling overheads remain practical for the long-running HPC and training workloads targeted in this paper.

\textbf{Runtime Overhead.} We first evaluate application-level performance loss (measured by runtime) across all systems. Several applications, such as \texttt{GPT2}, \texttt{pot3d}, and \texttt{ResNet101} on H100, exhibit noticeable performance loss. This is expected because EcoSched intentionally makes energy--performance trade-off decisions (e.g., scaling down GPU count) to unlock energy savings and additional co-scheduling opportunities. More generally, moderate per-application slowdown is often the intended mechanism by which EcoSched frees GPUs for concurrent execution and improves energy efficiency. NUMA-aware resource partitioning is also not perfect: some applications still incur small performance loss (e.g., 5\%) when a single application occupies three GPUs and one GPU resides in the other NUMA domain. In this case, accesses to the remote GPU may involve cross-NUMA uncore and memory-path resources, introducing additional overhead. Finally, \texttt{miniweather} on V100 shows a larger performance loss (40\%) because EcoSched scales its GPU count from the performance-optimal value of four down to one; however, this choice yields a 20\% energy saving for its running compared with using all GPUs. Separately, EcoSched's score-based online action-selection overhead is negligible, incurring less than 0.5 ms of decision-making overhead in our experiments.

\textbf{Energy Overhead.} The main question for profiling overhead is whether its energy cost remains bounded and can be amortized in practice. The energy overhead of EcoSched's online profiling is small and bounded: on H100, the profiling energy for each application in our workload is below 70~kJ. This cost can be amortized in two ways. First, EcoSched can recover the profiling cost through reduced active energy from better performance--energy tradeoffs. For example, \texttt{gpt2} consumes 64~kJ during profiling. The fastest profiled configuration for \texttt{gpt2} uses 3 GPUs and draws total of 1287~W of active GPU power, while EcoSched selects 2 GPUs with total of 946~W of active GPU power. This reduces active GPU power by 341~W, amortizing the profiling cost after 3.13~min of execution. Second, EcoSched can amortize profiling by reducing idle-GPU energy through co-scheduling. For example, \texttt{vgg16} consumes 34.00~kJ during profiling and EcoSched selects 1 GPU on H100. When other co-running applications use the remaining idle GPUs, EcoSched avoids as much as $3 \times 70$~W of idle-GPU power, amortizing the profiling cost after 2.70~min of overlap. In practice, ML training workloads commonly run for hours, so these one-time profiling costs are quickly amortized.

\begin{figure}
    \centering
    \includegraphics[width=1\linewidth]{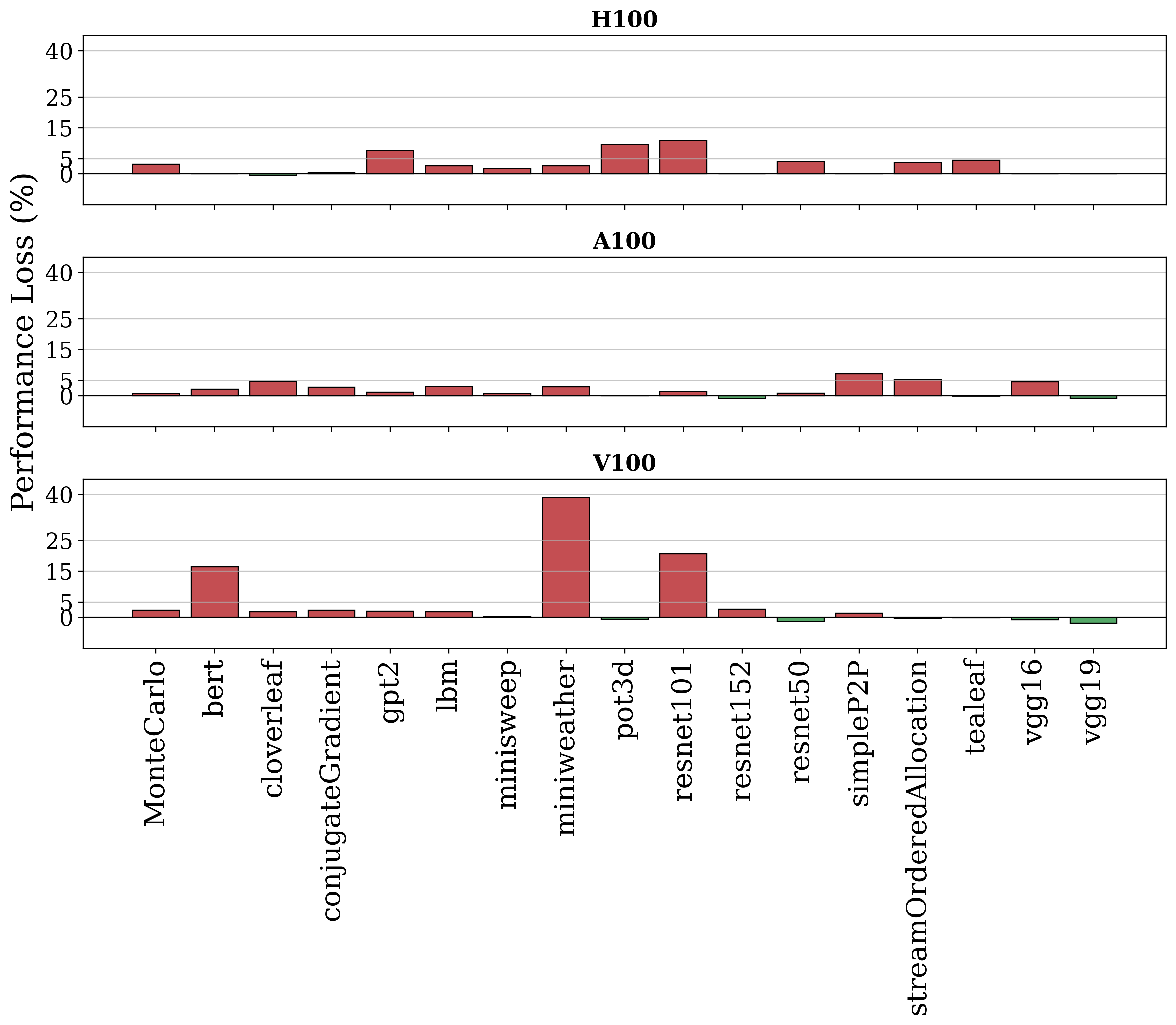}
    \caption{Per-application performance loss (measured by runtime) under EcoSched, relative to solo execution with the performance-optimal GPU count across all systems.}
    \label{fig:perf_loss}
    \vspace{-2em}
\end{figure}

\section{Discussion and Limitations}

This discussion clarifies the practical scope of EcoSched, the workload regimes it targets, how it fits into production scheduling environments, and the main dimensions along which the current prototype can be extended.

\textbf{Target workloads.} EcoSched is designed for long-running workloads whose execution time is large relative to the one-time profiling cost. In particular, the intended scope includes HPC scientific simulations and ML training workloads, where the brief online profiling overhead can be amortized over long executions and reused across repeated runs. By contrast, short-lived inference, serverless, or Function-as-a-Service (FaaS) workloads are outside the intended scope, because the profiling cost would represent too large a fraction of runtime.

\textbf{Scheduler integration.} EcoSched is not intended to replace the global batch scheduler. Instead, it can serve as a local node-level decision module that selects GPU counts and co-scheduling actions for a bounded set of candidate jobs. EcoSched operates on a bounded queue window rather than the full queue, which is a practical choice for online HPC scheduling. Because jobs arrive and complete continuously, production schedulers rarely know the future queue with certainty. A bounded window keeps the decision space manageable and supports fast re-scheduling at each scheduling event. EcoSched can therefore be integrated within an existing batch scheduler while the global scheduler continues to enforce policies such as priority, fairness, and backfilling. As the queue evolves, the window is refreshed and EcoSched is invoked again.

\textbf{Platform portability.} The current prototype is tied to NVIDIA GPUs because it relies on DCGM and NVML for telemetry and power measurement. However, the scheduling logic itself is not inherently vendor-specific. In principle, EcoSched can be ported to other GPU platforms if equivalent telemetry, power measurement, and GPU assignment controls are available, since the core design depends on lightweight online signals and controllable placement rather than on NVIDIA-specific scheduling assumptions.

\textbf{GPU sharing.} EcoSched intentionally focuses on whole-GPU allocation, which is the common granularity in current HPC batch scheduling. Mechanisms such as NVIDIA MIG and MPS are complementary to this scope, but incorporating them would expand the action space from GPU-count selection to joint selection of GPU partitions, sharing modes, and co-scheduling decisions. Such an extension would require additional modeling of interference, partition-specific performance, and energy behavior. These considerations clarify EcoSched’s role as a practical node-level scheduler for long-running multi-GPU HPC workloads, while leaving finer-grained sharing and broader platform support as natural future extensions.

\section{Related Work}

\textbf{Energy optimization in HPC.} A large body of work has studied energy optimization in HPC systems through mechanisms such as CPU/GPU DVFS, power capping, and power--performance modeling \cite{ramesh2019understanding,walker2018hardware,wallace2016application,bailey2015finding,freeh2005using,ge2007cpu,hsu2005power,lim2006adaptive,zheng2025minimizing,zheng2025coordinated}. These studies have shown that substantial energy savings can often be achieved by relaxing performance slightly or by tuning device-level operating points to match application behavior. However, most of this line of work optimizes the energy efficiency of a single application or a node-level operating point, rather than the scheduling decisions across a queued workload. As a result, these studies do not address the online scheduling problem considered here: jointly choosing GPU counts and co-running job sets across multiple queued applications while balancing energy efficiency and end-to-end throughput.

\textbf{GPU scheduling and co-scheduling in multi-GPU systems.} Prior work has studied scheduling and co-locating GPU applications in shared environments, especially in cloud systems \cite{peng2018optimus, zheng2019cynthia, xiao2018gandiva, gu2019tiresias, park2019accelerated, campos2017scaling,liu2023intelligent}. These schedulers improve utilization, fairness, or job completion time through techniques such as packing, preemption, migration, and elastic GPU allocation, and some jointly reason about allocation and placement. However, they typically rely on elastic execution models such as dynamic scaling, migration, or runtime reconfiguration, which differ fundamentally from HPC batch environments where jobs are launched with static resource allocations and runtime control is more limited. In HPC, prior node-sharing work has largely focused on CPU-only systems \cite{cpu_share_1,cpu_share_2,cpu_share_3,cpu_share_4,cpu_share_5,cpu_share_6,cpu_share_7}, while relatively few studies target multi-GPU nodes. Marble \cite{han2020marble} is the closest prior work in this setting and shows that profiling-guided co-location can improve utilization on multi-GPU HPC nodes. However, it focuses primarily on utilization, relies on comprehensive offline profiling, and does not jointly exploit energy-aware GPU-count selection, lightweight online estimation, and NUMA-aware placement. As a result, prior shared-system and HPC co-scheduling work does not fully address the static-allocation, online joint scheduling problem considered here.

Taken together, these lines of work address related aspects of the problem but not their combination. Prior energy-aware control work does not consider online scheduling across queued applications, shared-system GPU schedulers typically rely on elastic execution models that do not match static HPC constraints, and prior multi-GPU co-scheduling work improves utilization without combining joint energy-aware GPU-count selection, lightweight online estimation, and NUMA-aware placement. EcoSched’s novelty lies in integrating these elements into a single practical design.

\section{Conclusion}
This paper presents EcoSched, an online energy-aware co-scheduler for multi-GPU systems that jointly selects GPU counts and co-running jobs. With lightweight online modeling and NUMA-aware scheduling, EcoSched makes practical decisions without exhaustive offline profiling. Across H100, A100, and V100 platforms, EcoSched improves energy and EDP with acceptable performance impact, while capturing much of the benefit of an offline oracle. Future work includes deeper integration with cluster schedulers, stronger interference-aware models (e.g., PCIe/NVLink contention), and improved portability and adaptation across hardware platforms.


\newpage

\bibliographystyle{ieeetr}
\bibliography{bib/ipdps}

\end{document}